\documentclass[twocolumn, prx, floatfix,superscriptaddress,nobibnotes]{revtex4-1}

\usepackage{amsmath}
\usepackage{amssymb}
\usepackage{graphicx}
\usepackage{bm}
\usepackage{relsize}
\usepackage{color}
\usepackage{float}

\begin{document}

\title{Nonlocality of Majorana Modes in Hybrid Nanowires}

\author{M.-T.~Deng}
\affiliation{Center for Quantum Devices and Station Q Copenhagen, Niels Bohr Institute, University of Copenhagen, Copenhagen, Denmark}
\author{S.~Vaitiek\.{e}nas}
\affiliation{Center for Quantum Devices and Station Q Copenhagen, Niels Bohr Institute, University of Copenhagen, Copenhagen, Denmark}
\author{E.~Prada}
\affiliation{Departamento de F\'{i}sica de la Materia Condensada, Condensed Matter Physics Center (IFIMAC) and Instituto Nicol\'{a}s Cabrera, Universidad Aut\'{o}noma de Madrid, E-28049 Madrid, Spain }
\author{P.~San-Jose}
\affiliation{Materials Science Factory, Instituto de Ciencia de Materiales de Madrid, Consejo Superior de Investigaciones Cient\'{i}ficas (ICMM-CSIC), Sor Juana In\'{e}s de la Cruz 3, 28049 Madrid, Spain}
\author{J.~Nyg\aa rd}
\affiliation{Center for Quantum Devices and Station Q Copenhagen, Niels Bohr Institute, University of Copenhagen, Copenhagen, Denmark}
\author{P.~Krogstrup}
\affiliation{Center for Quantum Devices and Station Q Copenhagen, Niels Bohr Institute, University of Copenhagen, Copenhagen, Denmark}
\author{R.~Aguado}
\email{raguado@icmm.csic.es}
\affiliation{Materials Science Factory, Instituto de Ciencia de Materiales de Madrid, Consejo Superior de Investigaciones Cient\'{i}ficas (ICMM-CSIC), Sor Juana In\'{e}s de la Cruz 3, 28049 Madrid, Spain}
\author{C.~M.~Marcus}
\email{marcus@nbi.dk}
\affiliation{Center for Quantum Devices and Station Q Copenhagen, Niels Bohr Institute, University of Copenhagen, Copenhagen, Denmark}

\date{\today}

\begin{abstract}
{Spatial separation of Majorana zero modes distinguishes trivial from topological midgap states and is key to topological protection in quantum computing applications. Although signatures of Majorana zero modes in tunneling spectroscopy have been reported in numerous studies, a quantitative measure of the degree of separation, or nonlocality, of the emergent zero modes has not been reported. Here, we present results of an experimental study of nonlocality of emergent zero modes in superconductor-semiconductor hybrid nanowire devices. The approach takes advantage of recent theory showing that nonlocality can be measured from splitting due to hybridization of the zero mode in resonance with a quantum dot state at one end of the nanowire. From these splittings as well as anticrossing of the dot states, measured for even and odd occupied quantum dot states, we extract both the degree of nonlocality of the emergent zero mode, as well as the spin canting angles of the nonlocal zero mode. Depending on the device measured, we obtain either a moderate degree of nonlocality, suggesting a partially separated Andreev subgap state, or a highly nonlocal state consistent with a well-developed Majorana mode.}
\end{abstract}

\maketitle
\section{Introduction}
Emergent Majorana bound states (MBSs) in topological superconductors~\cite{Kitaev2000, Read2000} appear capable of providing a naturally fault-tolerant basis for quantum computing \cite{Nayak2008, DasSarmaNPJ2015}. The so-called topological protection is based on  the separation, or nonlocality, of MBSs, which makes Majorana qubits immune to decoherence by a local disturbance (though not immune to quasiparticle poisoning \cite{Goldstein2011, Budich2012}). Following early proposals to realize topologically protected MBSs~\cite{Moore1991, Fu2008}, hybrid superconductor-semiconductor nanowires~\cite{Lutchyn2010, Oreg2010} have emerged as one of the leading platforms due to straightforward nano-fabrication and experimental control~\cite{Lutchyn2017, Aguado2017}. 

\begin{figure}[b]
\centering \includegraphics[width=7 cm]{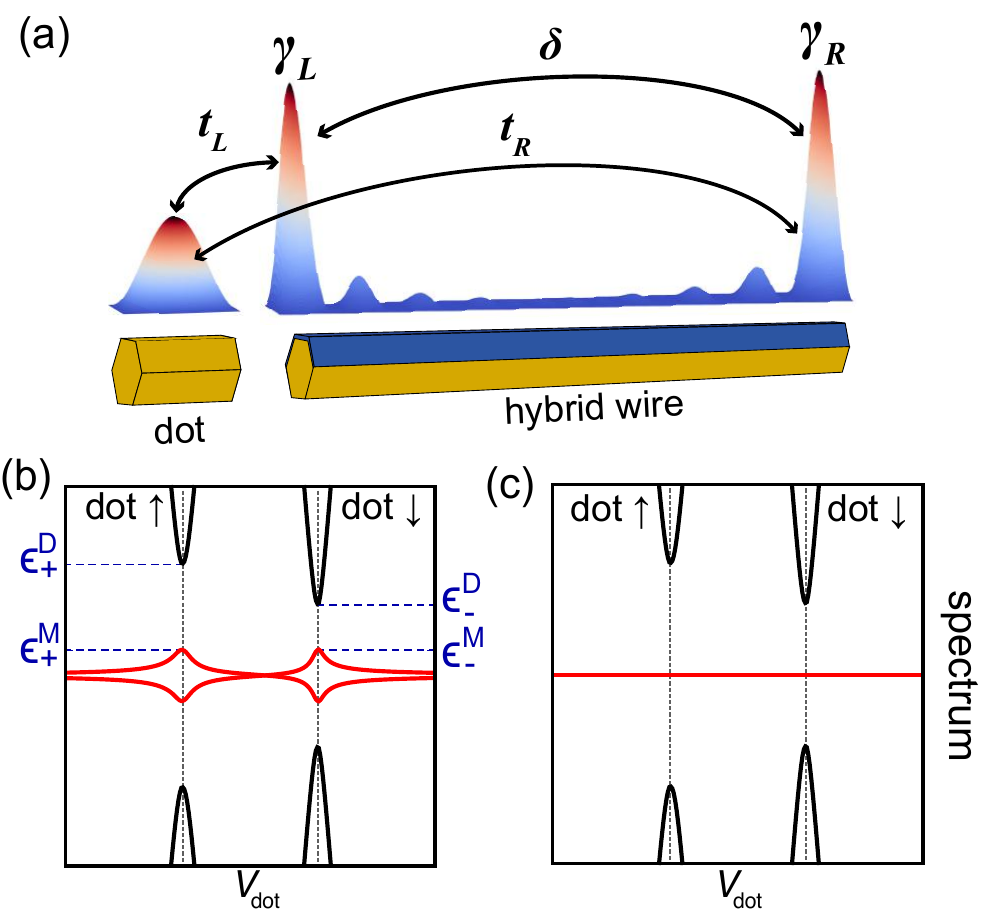}
\caption{\label{Fig1} (a) Schematic of Majorana bound states coupled to an end dot state. Two Majorana components, $\gamma_{L,R}$, are hybridized by $\delta$ due to direct interaction and further split by couplings $t_{L,R}$ to the dot. (b,c) The low-energy spectrum as a function of dot gate voltage $V_{\rm dot}$ shows anticrossings as dot states (black) align with (b) overlapping or (c) separated Majorana modes in the wire. Energies $\epsilon^{\rm M(D)}_{\pm}$ Majorana (dot) states are defined in the text.}
\end{figure}

In hybrid nanowires a few microns in length, MBSs evolve from Andreev bound states (ABSs), which coalesce under an increasing Zeeman field to zero energy in the presence of spin-orbit coupling. Theoretically, the degree of wave function nonlocality similarly evolves from an ABS with highly overlapping wave function components in the Majorana basis to two MBSs with spatially separated support at the wire ends (Fig.~\ref{Fig1}a). This spatial separation renders the Majorana qubit insensitive to local electromagnetic perturbations from the environment. Therefore, beyond observing zero-bias peaks~\cite{Mourik2012, Deng2012, Das2012, Churchill2013, Perge2014, Albrecht2016, Deng2016, Suominen2017, Nichele2017, Chen2017, Zhang2017}, the overlap of the two Majorana components, or degree of nonlocality, provides a metric of topological protection, or quality factor, for qubits realized through zero-energy modes. Proposed approaches to detect nonlocality include, for instance, gate dependence \cite{Liu2017}, quantum correlation \cite{Moore2017,Li2015}, and interferometry \cite{Sau2015, Rubbert2016, Hell2017}. However, these have not been demonstrated experimentally to date.

Here, we report an experimental investigation of zero mode nonlocality, using recent theory that shows how, under reasonable model assumptions, a local probe at one end of the wire allows the degree of nonlocality to be detected and quantified \cite{Prada2017,Clarke2017}. The approach exploits the insensitivity of nonlocal MBSs to local interaction, the feature ultimately responsible for topological protection. Specifically, fully separated MBSs remain at zero energy when perturbed by a local probe; splitting in response to a local probe indicates a finite Majorana overlap. A naturally occurring quantum dot at the end of a semiconductor-superconductor hybrid nanowire \cite{Deng2016} was tuned across a resonance with the MBS and the resulting splitting of the zero-energy state was measured by tunneling spectroscopy [Fig.~\ref{Fig1}(a)].  In addition, anticrossings between wire midgap states and consecutive dot states yield information about spin structure of the midgap state \cite{Prada2017}.
Relevant features in the data are schematically illustrated in Fig.~\ref{Fig1}(b,c), which contrast consecutive dot-wire anticrossings for overlapping versus separated MBSs.

\section{Approach to Measuring Nonlocality}

 In a topological nanowire of finite length $L_w$, the two Majoranas $\gamma_L$ and $\gamma_R$ at either end overlap and hybridize into fermionic subgap eigenstates $c_M=\gamma_L+i\gamma_R$ of energy $\delta$. The  wave functions $u^{(L,R)}_\sigma(x)$ of its two Majorana components $\gamma_{L,R}$ remain spatially separated, concentrated near the left ($x=0$) and right ($x=L_w$) ends of the wire, respectively. We define the degree of Majorana nonlocality as the overlap integral summed over spin index $\sigma$,
\begin{eqnarray}\label{omega}
\Omega&=&\sum_\sigma\int_0^{L_w} dx\left|u^{(L)}_\sigma(x)u^{(R)}_\sigma(x)\right|.
\end{eqnarray}
Highly nonlocal MBSs have $\Omega \rightarrow 0$, while the Majorana components of a conventional ABSs have $\Omega \rightarrow 1$. The absolute value in the definition Eq.~(\ref{omega}) means that a vanishing $\Omega$ implies {\em spatially separated} MBSs, and a nonlocal fermionic state $c_M$, see Fig.~\ref{Fig1}(a). Since the overlap integral is related to the charge of overlapping Majoranas \cite{ Ben-Shach2015,Pinning2017},  a nonlocal state with $\Omega=0$ has zero charge and therefore would remain unperturbed by local electromagnetic noise. This makes the magnitude of $\Omega$ of particular importance when using MBSs to implement a topological qubit. We note, in particular, that a fermionic zero mode with $\delta=0$ need not have $\Omega=0$ when written in the Majorana basis, as discussed below.

The approach employed here to experimentally measure the degree of Majorana nonlocality, i.e., the suppression of $\Omega$, was recently proposed in Refs.~\cite{Clarke2017, Prada2017}. It is based on detecting, through transport spectroscopy, the hybridization of a state in a quantum dot coupled to the left end of the nanowire as it comes into resonance with the Majorana mode.
Due to their spatial separation, the left and right Majoranas will couple to the dot state through different hopping amplitudes, $t_L$ and $t_R$, proportional to the respective Majorana wave functions at the position of the dot-wire barrier, as sketched in Fig.~\ref{Fig1}(a). This difference in coupling is reflected in the spectral lines of the coupled dot-nanowire system.
In the case of finite spatial overlap $\Omega$, the remote (right) Majorana acquires a finite coupling $t_R$ to the dot, and as a result the hybridized Majorana state $c_M$ becomes shifted in energy across resonances [Fig.~\ref{Fig1}(b)]~\cite{Deng2016, Deng2014}. Strictly nonlocal Majoranas, in contrast, have $\delta,t_R\approx 0$ and remain unperturbed at zero energy [Fig.~\ref{Fig1}(c)]. The energy shift can be used to extract the ratio $t_R/t_L$ and the spin canting angles $\theta_{L,R}$ of the two Majorana wave functions $u^{(L/R)}_\sigma$ at the contact \cite{Prada2017}. These spin orientations deviate from that of the dot states due to the spin-orbit coupling in the nanowire. 

A comparison to microscopic models indicates \cite{Prada2017} that in the topological regime, the dimensionless quantity $\eta = \sqrt{t_R/t_L}$ accurately estimates the overlap, $\Omega\approx \eta$, under generic conditions. Following similar arguments in a spinless context, a related quality factor of Majorana nonlocality, $q=1-\eta^{2}$, was introduced in Ref.~\cite{Clarke2017}. 
For $\delta\ll t_L,t_R$, the overlap estimate can be expressed \cite{Prada2017} in terms of energies $\epsilon_{\pm}^{M,D}$, which characterize consecutive anticrossings of dot and wire levels [see Fig.~\ref{Fig1}(b) and Appendix D],
\begin{equation}\label{eta}
\Omega^2\approx \eta^2=\frac{\epsilon_{-}^{M}}{\epsilon_{-}^{D}}\left|\frac{\sin\frac{1}{2}\theta_L}{\sin\frac{1}{2}\theta_R}\right|=\frac{\epsilon_{+}^{M}}{\epsilon_{+}^{D}}\left|\frac{\cos\frac{1}{2}\theta_L}{\cos\frac{1}{2}\theta_R}\right| \leq 1.
\end{equation}
The ratios of consecutive anticrossing energies can then be directly related to spin canting angles,  $\epsilon_{-}^{D}/\epsilon_{+}^{D}=\left|\tan\frac{1}{2}\theta_L\right|$ and $\epsilon_{-}^{M}/\epsilon_{+}^{M}=\left|\tan\frac{1}{2}\theta_R\right|$.
Fitting to this model for a given coupled dot-MBS crossing spectrum gives a unique combination of $t_{L,R}$ and $\theta_{L,R}$ in the range $0<\theta_{L,R}<\pi$ and $0<t_{R}<t_{L}$.

\begin{figure*}[t]
\centering \includegraphics[width=15 cm]{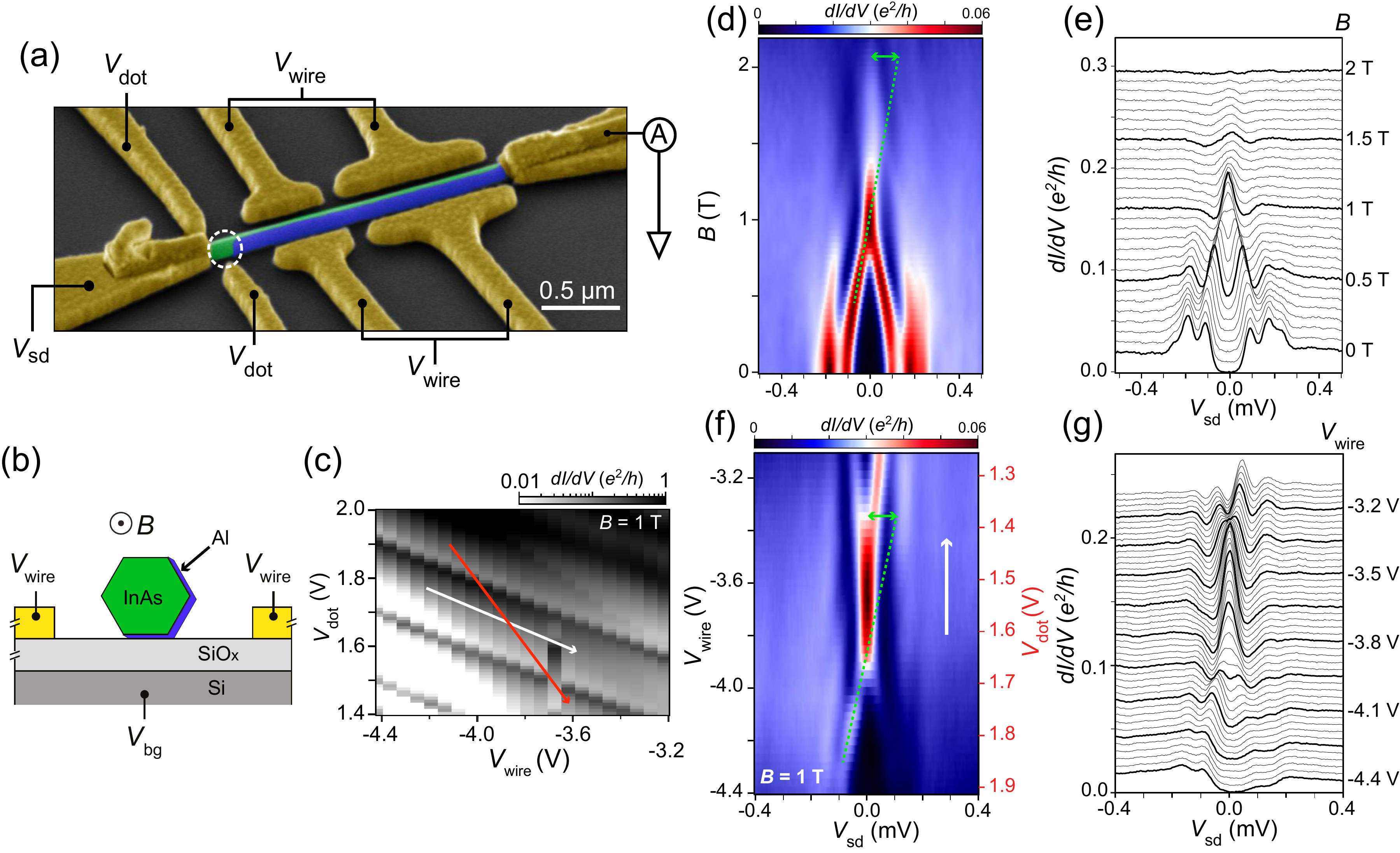}
\caption{\label{Fig2} (a) False-color micrograph of device A with end quantum dot (dashed circle). (b) Schematic cross section of device, with InAs core (green) and epitaxial Al (blue) on three facets. Magnetic field $B$ is applied parallel to the nanowire. (c) Conductance in $V_{\rm dot}$-$V_{\rm wire}$ plane at $B = 1$~T and $V_{\rm bg}=-2$~V. Arrows indicate isopotential directions for dot (white) and wire (red). (d) Color scale plot of differential conductance, measured at $V_{\rm dot}$ = 1.41~V, $V_{\rm wire}$ = -3.68~V and $V_{\rm bg}$ = -2~V, as a function of $V_{\rm sd}$ and $B$.  A zero-bias conductance peak emerges around $B \sim 0.8$~T and persists to $\sim 2$~T. In order to show the robustness of the zero-bias peak at the energy scales corresponding to Zeeman splitting, a green dashed line is extended from the merging ABS with the effective g-factor as its slope. (e) The same as (d), but on a line cut plot. (f) Conductance measured at $B$ = 1~T as a function of $V_{\rm sd}$ and the $V_{\rm dot}$-$V_{\rm wire}$ combined voltage following a wire isopotential [white arrow in (c)]. The zero-bias conductance peak persists over the range $V_{\rm wire} = -3.9$~V to $-3.3$~V. The slope of the extended green dashed line is fit from the gate lever arm. (g) The same as (f), but on a line cut plot. All panels are for device A.}
\end{figure*}

\section{Device and Experiment}
The devices studied are based on InAs nanowires with a 7--10~nm epitaxial Al layer on three facets of the wire, grown by molecular beam epitaxy \cite{Krogstrup2015}. Previous studies on similar nanowires showed a hard induced superconducting gap \cite{Chang2015}, with critical magnetic field along the wire axis exceeding 2~T~\cite{Deng2016}. The Al shell was etched on one end of the wire, leaving a bare InAs segment. Ti/Au (5/100~nm) ohmic contacts were deposited on both ends, forming a $\sim$150~nm bare InAs segment and a 2~$\mu$m InAs/Al segment between the contacts. Electrostatic control of wire and barrier density was provided by side gates and a global back gate, as shown in  Figs.~\ref{Fig1}(c) and (d). Three devices, denoted A-C, were measured, showing similar general behavior, though with different degrees of nonlocality. Measurements were carried out using standard ac lock-in methods in a dilution refrigerator with a three-axis vector magnet.
\begin{figure*}[ht]
\centering \includegraphics[width=14cm]{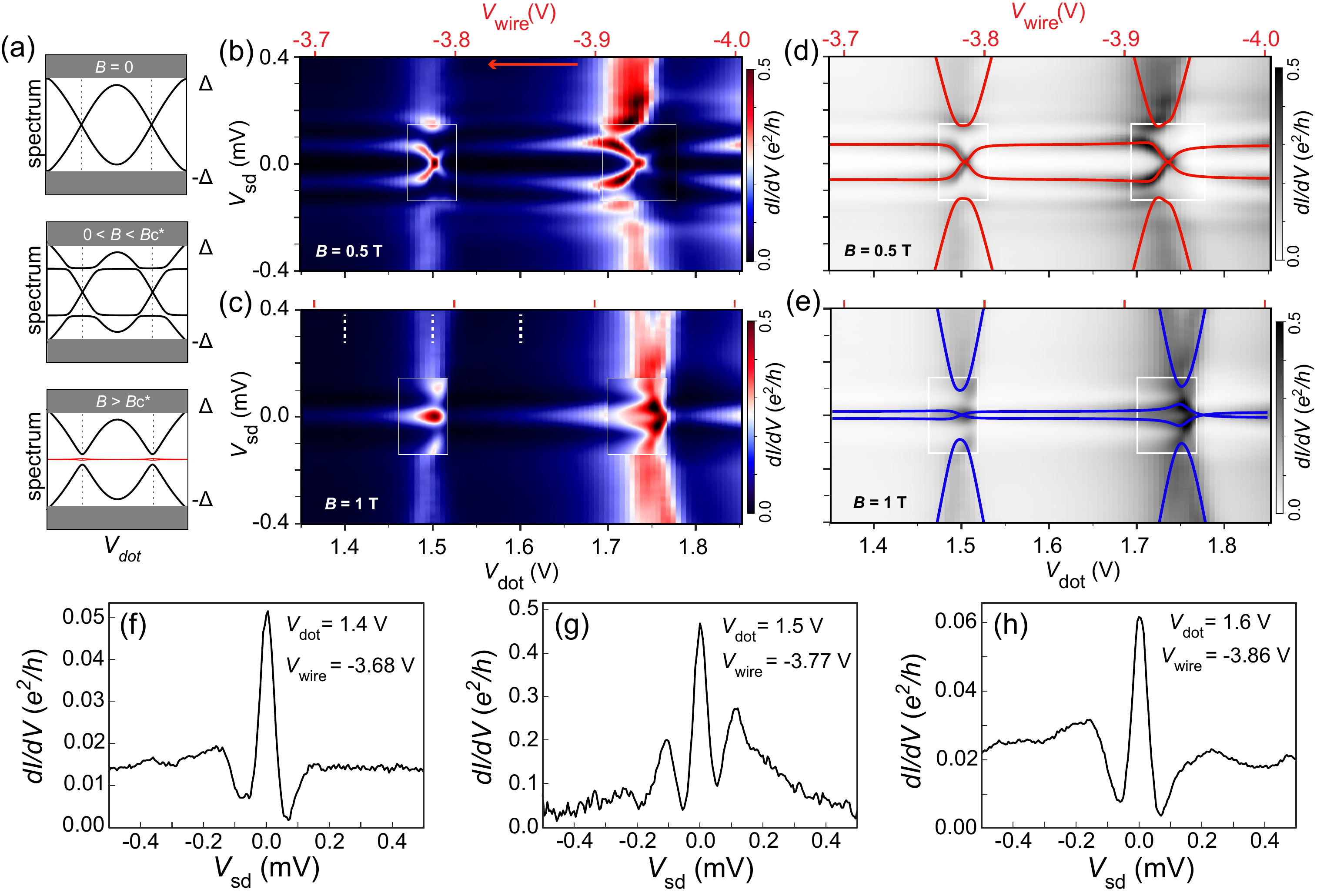}
\caption{\label{Fig3} (a) Theoretical regimes of dot-wire interactions. Trivial regime at zero magnetic field, with ABSs in the dot and all wire states outside the diagram (top panel); trivial regime with applied magnetic field that lowers one wire state into the gap (middle panel); topological regime, with dot states interacting with zero-energy wire state (bottom panel). (b) Differential conductance at $B = 0.5$~T for gate sweep along the wire-isopotential direction [red arrow in Fig.~\ref{Fig1}(e)]. (c) The same as (b), but at $B = 1$~T. Subgap wire state remains close to zero away from resonance with dot and splits when on resonance, reflecting the locality of the wire state (boxed regions measured at higher resolution). Inequivalence of consecutive resonances reflects spin interaction. (d,e) Model fits for (b,c) (red and blue lines), which are consistent with partially localized midgap state in wire. (f-h) Line cut plots taken from panel (c), at $V_{\rm dot}=1.4$~V, 1.5~V and 1.6~V, respectively. All panels are for device A.}
\end{figure*}
A quantum dot typically forms in the bare InAs segment at the end of the wire [dashed circle in Fig.~\ref{Fig2}(a)], presumably due to density of state gradients at the edges of the Al shell and metallic lead and possibly disorder. Details of the dot can be seen in Appendix A. Occupancy of the end quantum dot is tuned by the voltage $V_{\rm dot}$ on the gates close to the dot, while $V_{\rm wire}$ is used to tune density in the wire.  Voltage $V_{\rm bg}$ on a global back gate change both the dot level and the wire density. To separately tune the dot level or the wire chemical potential, $V_{\rm dot}$ and $V_{\rm wire}$ are changed in a compensatory way~\cite{Deng2016}. Figure~\ref{Fig2}(c) shows a 2D plot of $V_{\rm dot}$ and $V_{\rm wire}$ with high-conductance stripes indicating quantum dot resonances.  Compensated sweep directions for dot isopotentials (white arrow direction) and wire isopotentials (red arrow direction) are also indicated in subsequent figures. The wire isopotential direction is determined by the onset of wire state appearance for a large range including several dot Coulomb valleys [see Fig.~\ref{Fig5}(d) in Appendix A]. To find zero-bias peaks, the end-dot is tuned to a Coulomb-blockade valley and used as a co-tunneling spectrometer, with gates swept parallel to the white arrow~\cite{Deng2016}. To pass through dot resonances at fixed wire density, gates are swept parallel to the red arrow as described in detail below.

Tunneling spectroscopy reveals an induced superconducting gap of $\sim $200~$\mu$eV at zero field, along with a pair of subgap ABSs at $\sim \pm100$~$\mu$eV [Fig.~\ref{Fig2}(d)]. With increasing field, the ABSs split and move to lower energy with an effective \emph{g}-factor of $\sim$3.5, merging into a zero-bias peak around $B$ = 0.8~T that persists up to $B = 2$~T [Fig.~\ref{Fig2}(d,e)]. In gate voltage dependence measurements, the zero-bias peak extends from $V_{\rm wire}=-3.95$~V to $V_{\rm wire}=-3.25$~V along the dot isopotential [Figs.~\ref{Fig2}(f) and (g)]. To take into account SC-SM hybridization~\cite{MBSgfactor, Antipov2018, Mikkelsen2018, Reeg2018}, we have also estimated the robustness of the zero-bias peak at the energy scales corresponding to Zeeman splittings or potential variations. By using the extracted \emph{g}-factor and the gate lever arm, the zero-bias peaks in Figs.~\ref{Fig2}(d) and (f)  extend for $\sim$120~$\mu$eV  [green dashed lines in Figs.~\ref{Fig2}(d)] and $\sim$100~$\mu$eV [green dashed lines in Figs.~\ref{Fig2}(f)], both sufficiently exceeding the peak broadening ($\sim$45~$\mu$eV).

The theoretical regimes of behavior for the dot-wire system are illustrated schematically in Fig.~\ref{Fig3}(a). At zero magnetic field and with no wire states below the dot charging energy, ABSs in the dot form loops (top panel), the smaller for odd dot occupancies, as observed previously~\cite{Pillet2010, Mason2011, Chang2013, Lee2013}. As subgap wire states split and move toward zero energy with applied field, dot and wire states anticross (middle panel). Finally, in the topological regime, wire states coalesce to a zero-energy state (red line),  which can split when interacting with dot states, depending on the separation of the two MBSs (bottom panel).

To apply the above method to the midgap state in device A, the gate voltages have been swept along a wire isopotential, passing through consecutive resonances of the end dot. Figure~\ref{Fig3}(b) shows interaction of dot states and wire ABSs at $B = 0.5$~T, plotting conductance along with model spectra, as described below.  At $B=1$~T, the resonance between the zero mode and the dot level at $V_{\rm dot}\sim 1.75$~V appears as a splitting of the zero-bias peak into a characteristic diamond shape [Fig.~\ref{Fig3}(c)]. In contrast, the zero-bias peak at $V_{\rm dot} \sim 1.5$~V does not show any detectable splitting. Throughout the whole $V_{\rm dot}$ gate scan, the near-zero subgap state remains disconnected from the visible gap edge or other subgap states.  

Using Eq.~(\ref{eta}), the fittings for data in Figs.~\ref{Fig3}(b) and (c) are illustrated in Fig.~\ref{Fig3}(d) (red lines) and Fig.~\ref{Fig3}(e) (blue lines), respectively, with $t_{L,R}$, $\delta$ and $\theta_{L,R}$ as free parameters. The fitting yields $\eta\sim 0.5$, corresponding to $q\sim 0.75$, and spin-canting angles, $\theta_L\sim 1.7$, $\theta_R\sim 3.0$ (in radians). This provides an illustrative example of a nearly unsplit zero energy midgap state, yet with only a moderate degree of nonlocality.  

The spectrum also differs markedly between consecutive resonances, consistent with theory for spin-dependent anticrossings \cite{Prada2017}.  The fit of the model (Section II and Appendix D) shows  excellent agreement with experimental data across both resonances. Additionally, the different visibility of the zero energy and excited states is the result of the dot-wire spin alignment for consecutive resonances, which determines the degree of wave function leakage into the dot, also consistent with theoretical expectations \cite{Chevallier2018}. 

\begin{figure}
\centering \includegraphics[width=8.5cm]{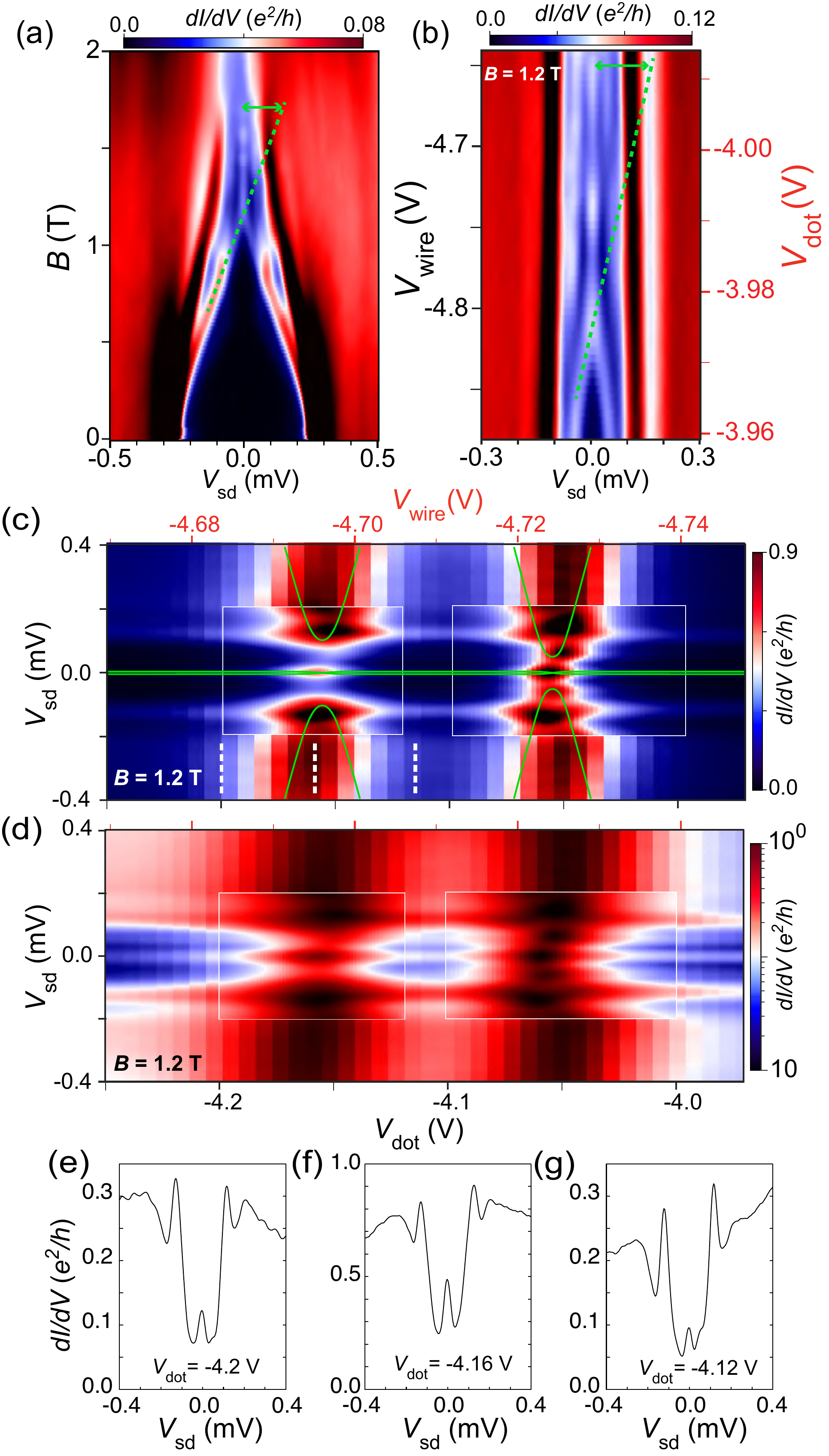}
\caption{\label{Fig4} (a) Differential conductance as a function of $V_{\rm sd}$ and magnetic field $B$ at $V_{\rm bg}$~=~0, $V_{\rm dot}$~=~-3.9~V and $V_{\rm wire}$~=~-5.0~V. Gap reduction and reopening around $B$~=~1~T is accompanied by appearance of zero-bias peak from $B\sim$1.1~T to 1.7~T. (b) Differential conductance along dot-isopotential direction at $B = 1.2$~T. Zero-bias peak persists for large gate-voltage range. (c) Dot-wire resonances along a wire-isopotential. The zero-bias peak persists without splitting through resonances, while it anticrosses with the dot states. Model fits (green) consistent with nonlocal MBSs in the wire. (d) Same as (c), but in logarithmic color scale. Unsplit zero-bias peak is visible through the entire gate range. (e)-(g) Line cut plots taken at white dashed lines in panel (c). All panels are for device B.}
\end{figure}

Applying the same measurements for device B, a narrow and stable zero-bias peak is found to span the range $B\sim 1.1-1.7$~T (corresponding to $\sim$160~$\mu$eV Zeeman splitting, the peak broadening is $\sim$20~$\mu$eV), with weak oscillations in height and width above 1.4~T [Fig.~\ref{Fig4}(a)]. The emergence of the zero-bias peak was accompanied by a near-closing and reopening of the gap to excited states. The gate sweep in Fig.~\ref{Fig4}(b) shows the stability of the midgap state as a function of the wire potential, which demonstrates that the zero-bias peak is robust in a range of gate potential variations of $\sim$160~$\mu$eV.

For gate sweeps along the wire isopotential [Figs.~\ref{Fig4}(c) and (d)], the zero-bias peak remains locked at zero bias, insensitive to the dot state through wire-dot resonances---in contrast to device A---while the dot state shows large anticrossings with the zero-energy wire state through wire-dot resonances. Non-splitting of the zero-bias peak across a pair of resonances [Fig.~\ref{Fig4}(c)] is not fine-tuned, and is found to be robust to changes in magnetic field [Appendix B, Fig.~\ref{Fig6}(c)] and wire potential [Appendix B, Fig.~\ref{Fig6}(d)]. Model fits yield $\eta < 0.17$, which corresponds to $q > 0.97$, and spin canting angles $\theta_L\sim0.9$, $\theta_R\sim1.6$, indictating that the midgap states in device B are consistent with highly nonlocal MBSs.  The interpretation of a large nonlocality for device B is further supported by the small residual gap evident near the phase crossover point [Fig.~\ref{Fig4}(a)]~\cite{Deng2016, Mishmash2017}, a feature that is absent in device A [Fig.~\ref{Fig2}(d)]. 

The nonlocality can vary with devices, presumably due to defects induced during fabrication (like wire transferring). It also strongly depends on gate tuning, which can change potential profile and/or effective Fermi wave vector and results in change of nonlocality. In Appendix B, measurements have been also performed at different gate configurations for device B, with a different nonlocality factor.

Data for a third device, denoted C, are shown in Appendix C. These data indicate a significantly overlapping midgap states, presumably associated with a trivial ABS. 

\section{Summary and Discussion}

The key property of Majorana zero modes to serve as the basis for topological quantum computing is their insensitivity to local disturbance or measurement arising from their spatial separation, or nonlocality. That same feature, insentivity to a local probe, has been applied in this study to quantify the separation of MBSs, taking advantage of recent theoretical developments concerning how the Majorana modes individually couple to a quantum dot at one end of the nanowire. We found that for two devices (devices A and C), the near-zero-bias conductance peak corresponds to an intermediate degree of nonlocality, while in another (device B), the zero-bias peak remained at zero when passing through consecutive dot resonances, characteristic of highly nonlocal MBSs. Fits to theory yield bounds on a measure of locality, $\eta < 0.17$, and quality factor $q > 0.97$, as well as measurement of the spin canting angles of the zero-mode spin texture. 

Arguably more important than the values of nonlocality that were obtained is the experimental demonstration of a new method for estimating nonlocality and quality factor that can be applied to existing nanowire devices. The measurement of spin canting angles using consecutive resonances adds a further identifier to the Majorana modes. It would be interesting to correlate spin canting angles measured using two end dots at opposite ends of a nanowire, presumably allowing one to establish that the same two Majoranas are being measured. That study will be pursued in future experiments.

\section*{Acknowledgements}
We thank David Clarke, Karsten Flensberg and Martin Leijnse for valuable discussions, and Claus S{\o}rensen and Shivendra Upadhyay for contributions to material growth and device fabrication. Research supported by Microsoft, the Danish National Research Foundation, the European Commission, and the Spanish Ministry of Economy and Competitiveness through Grants FIS2015-65706-P, FIS2015-64654-P, FIS2016-80434-P (AEI/FEDER, EU), the Ram\'on y Cajal programme Grant RYC-2011-09345, and the Mar\'ia de Maeztu Programme for Units of Excellence in R\&D (MDM-2014-0377). C.M.M. acknowledges support from the Villum Foundation. M.-T.D. acknowledges support from State Key Laboratory of High Performance Computing, China. 

M.-T. D.~and S.V.~contributed equally to this work.

\appendix 

\section{Supporting data for device A}

In this section, we present some basic analysis on the quantum dot in device A.

\begin{figure}[h]
\centering \includegraphics[width=8 cm]{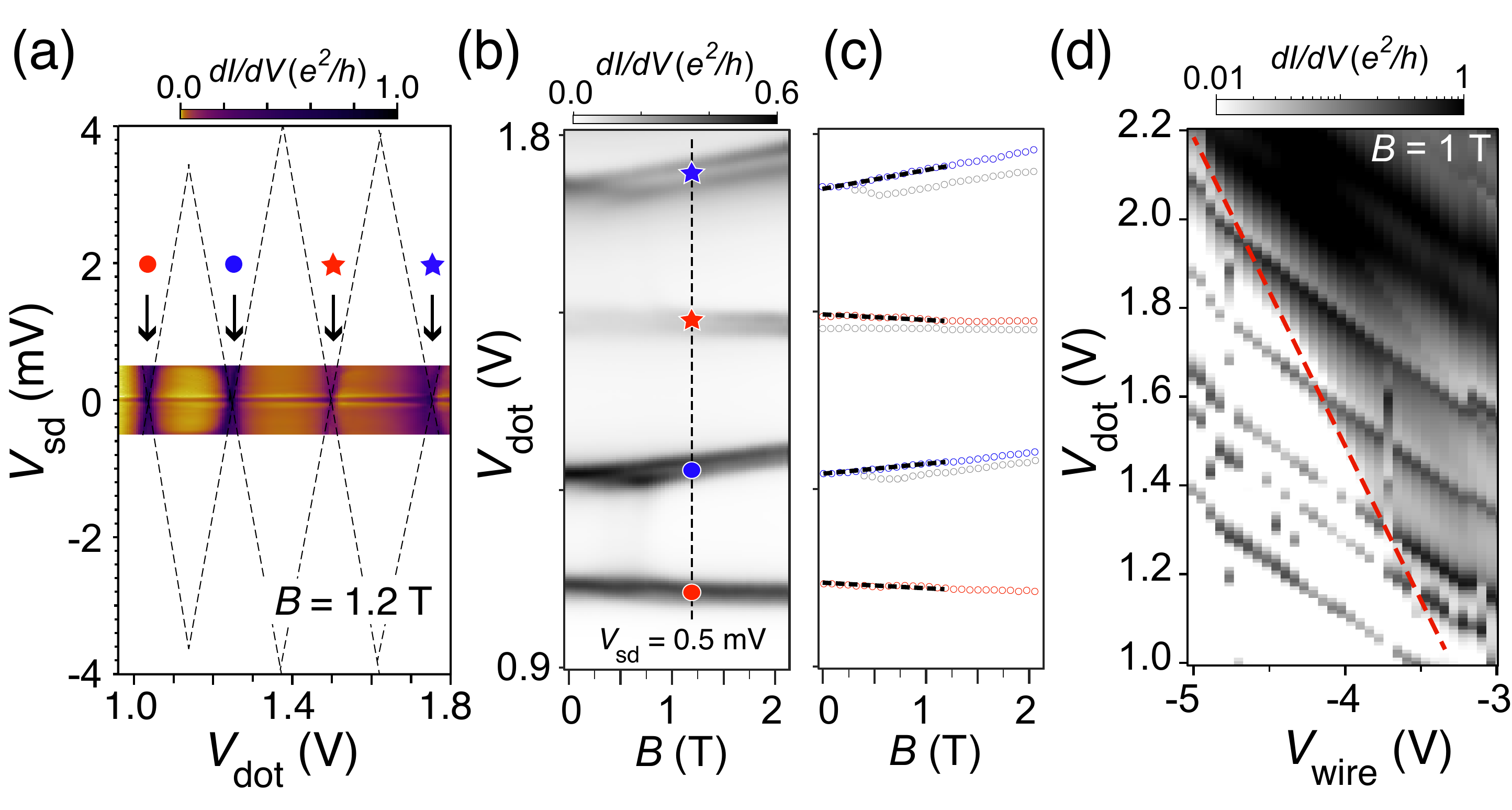}
\caption{\label{Fig5} (a) Differential conductance measured for device A, as a function of source-drain bias voltage and the $V_{\rm dot}$-$V_{\rm wire}$ combined voltage along the wire-isopotential direction [red arrow in Fig.~\ref{Fig2}(c)]. The measurements are performed at $B = 1.2$ T.  (b) The quantum dot level evolution in magnetic field, measured at $V_{\rm sd}=0.5$ mV. Corresponding levels are labeled with the same mark as in panel (a). (c) Extracted peak positions in panel (b). (d) Large scale view of Fig.~\ref{Fig2}(c). The red dashed line indicates the onset of wire state appearance. }
\end{figure}

As shown in Fig.~\ref{Fig5}(a), by extending the Coulomb blockade diamond edges at low energy, we can roughly estimate the addition energy of the dot to be 3$\sim$4 meV at $B = 1.2$~T, and the lever arm of the combined $V_{\rm dot}$-$V_{\rm wire}$ along the wire-isopotential direction is about 0.016. We also measured magnetic field dependence of the dot levels [Fig.~\ref{Fig5} (b)]. At zero field, four dot levels are separated by $\sim$3.6~meV, $\sim$4.2~meV and $\sim$3.0~meV addition energy in succession, which is consistent with the spin 0 and spin 1/2 alternating filling picture. As magnetic field increases, the up two levels evolve towards opposite gate voltages, which can be attributed to Zeeman splitting of the InAs end dot. Fitting the splitting slopes in low field regime [Fig.~\ref{Fig5}(c)], an effective g-factor around 10 can be deduced. The bottom two levels show similar splitting behavior with a g-factor about 7. Note the slitted peaks in Fig.~\ref{Fig5}(b) is because the access of excited state at $V_{\rm sd} = 0.5$~mT.

At base temperature, the dot resonant levels are much broader than the zero-energy mode in the wire. This is because the dot has to be strongly coupled to the superconductor to achieve large level anticrossing, and therefore the dot levels are lifetime broadening dominated.

\section{Supporting data for device B}

\begin{figure}[H]
\centering \includegraphics[width=8cm]{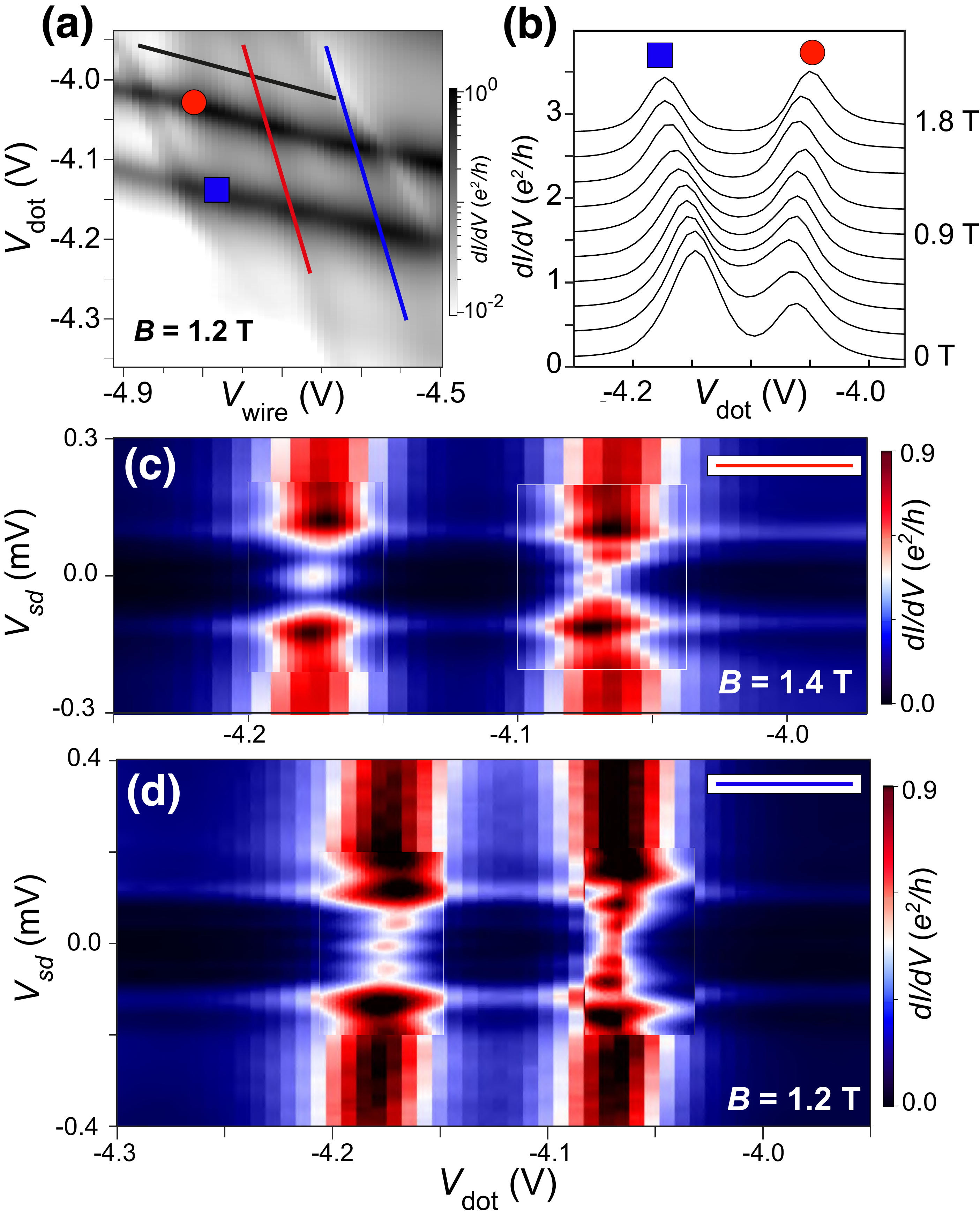}
\caption{\label{Fig6} (a) Logarithmic color scale plot of differential conductance for device B, measured at $V_{\rm sd} = 0$, $B=1.2$~T, as a function of $V_{\rm dot}$ and $V_{\rm wire}$. The black and red lines denote dot-isopotential and wire-isopotential gate sweeping directions, respectively. Figure~\ref{Fig4}(b) of the main paper is taken along the black line, while Fig.~\ref{Fig4}(c) is taken along the red line. (b) Magnetic field evolution of the quantum dot levels shown in (a) (solid circle and box). The Zeeman splitting induced level shift is evident. (c) Similar to Fig.~\ref{Fig4}(c), but measured at $B = 1.4$~T (along the red line). (d), Similar to Fig.~\ref{Fig4}(c), but measured along a different wire-isopotential direction (along the blue line), at $B = 1.2$~T. Same as Fig.~\ref{Fig4}(c), the ZBPs in (c) and (d) show the same robustness when they cross the dot levels, while the anticrossings between dot states and wire states remain large.}
\end{figure}
Here we present supporting data for device B. 

The conductance in $V_{\rm dot}$-$V_{\rm wire}$ plane clearly depicts the dot-levels (red dot and blue square). The evolution of the dot levels in magnetic field shows Zeeman splitting induced level shift in gate.

Measurements for device B that are similar to Fig.~\ref{Fig4}(c) have also been taken at different magnetic field [Fig.~\ref{Fig6}(c)] and along different slope [Fig.~\ref{Fig6}(d)]. Those measurements all show that the zero-bias peaks remain unsplit when they cross the consecutive dot levels with finite anticrossings between dot states and wire states. 

\begin{figure}
\centering \includegraphics[width=8.5 cm]{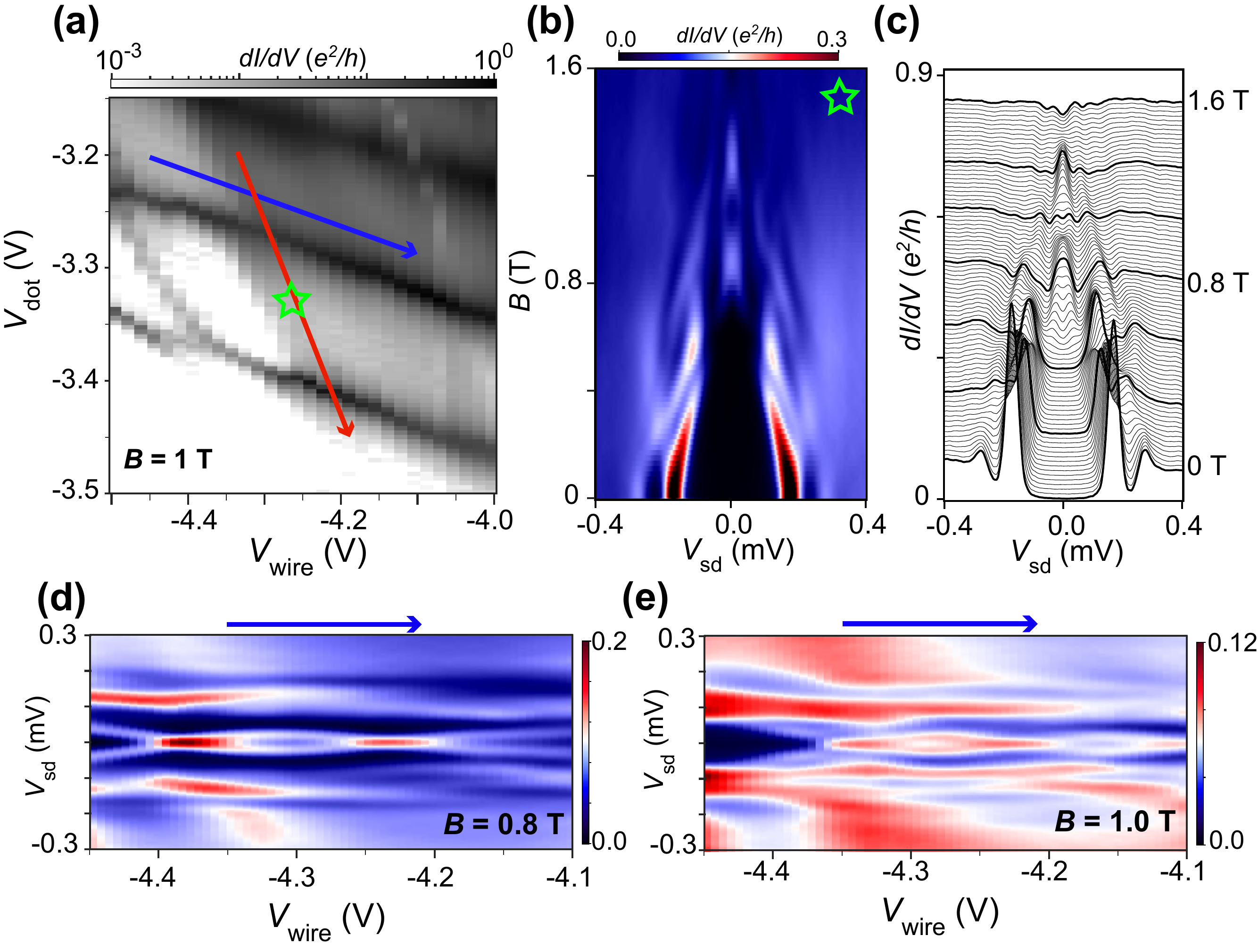}
\caption{\label{Fig7} (a) The gate map of $V_{\rm dot}$ and $V_{\rm wire}$, at $V_{\rm sd} = 0$, $V_{\rm bg} =-2$~V and $B = 1$~T. The blue and red lines represent the dot-isopotential and wire-isopotential directions, respectively. (b), Tunneling spectrum evolution in magnetic field, measured at the point indicated by the star in panel (a). A zero-bias peak emerges at $B =0.8$~T, and then it goes through a splitting-merging process. (c) A line-cut plot of panel (b). (d) and (e) Tunneling spectra along a dot-isopotential direction [the blue line in panel (a)] measured at $B=0.8$~T and $B$ = 1~T, respectively. Along the gate voltage, the sub-gap states show oscillatory splittings around zero bias voltage. }
\end{figure}

Topological phase strongly depends on the chemical potential of the wire. Here, we performed nonlocality measurements for device B again, but with a quite different gate configuration. The back gate voltage $V_{\rm bg}$ is tuned to -2 V, and the nanowire is supposed to be depleted more. At this back gate voltage, we have performed $V_{\rm dot}-V_{\rm wire}$ gate sweep and marked the dot-isopotential and wire-isopotential directions in Fig.~\ref{Fig7}(a). A tunneling spectrum in magnetic field is applied at the point marked by a star in Fig.~\ref{Fig7}(a), where an emergent zero-bias peak can be seen around $B=0.8$~T [Fig.~\ref{Fig7}(b) and (c)]. This zero-bias peak ``oscillates" at higher magnetic field and formed a loop structure. Gate sweep is also performed along a dot-isopotential direction [blue arrow in Fig.~\ref{Fig7}(a)]. The oscillatory splitting of zero-bias peak is evident to see both at $B = 0.8$~T and $B = 1.0$~T [Fig.~\ref{Fig7}(d) and (e)].

\begin{figure}
\centering \includegraphics[width=8.5 cm]{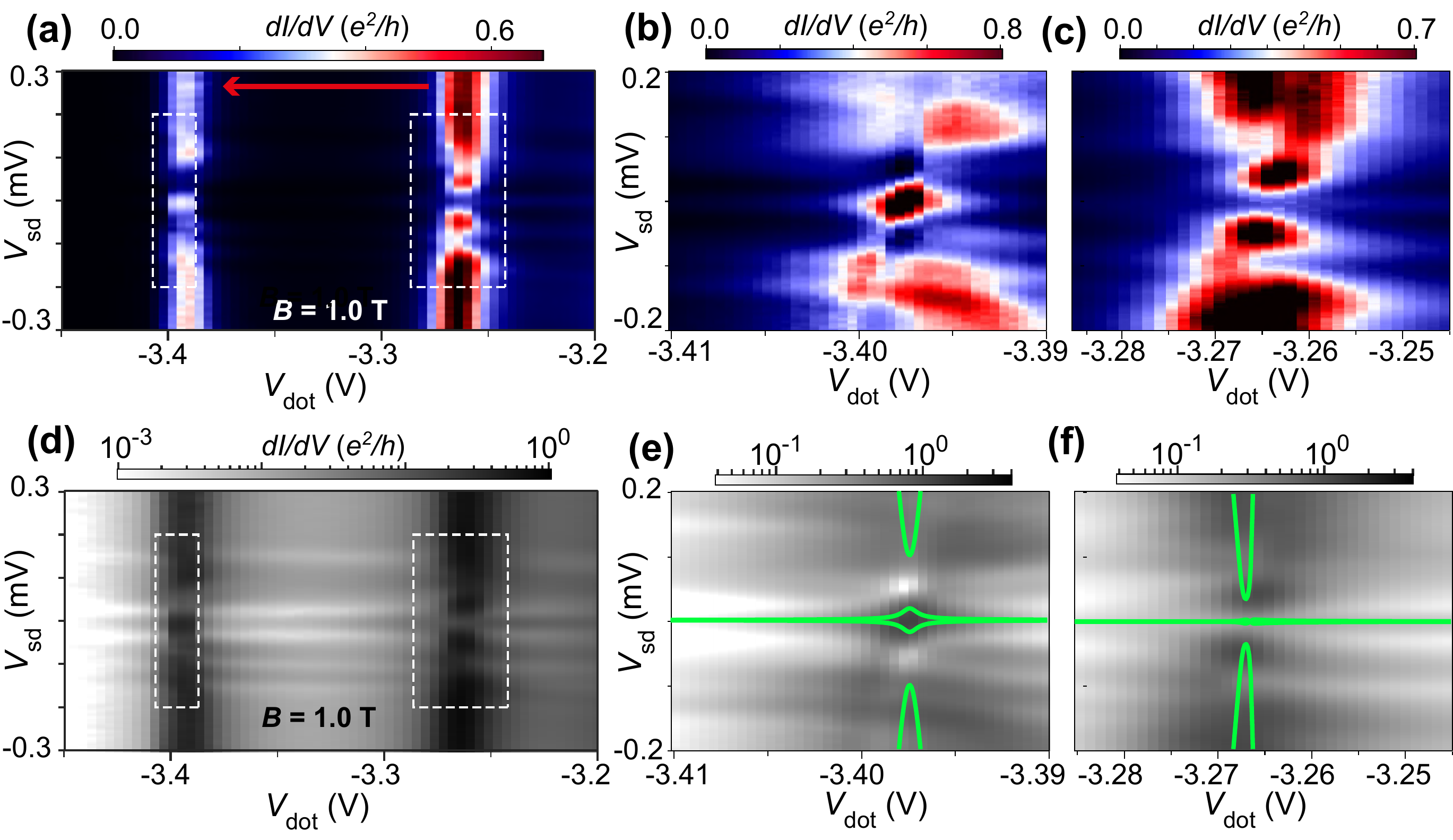}
\caption{\label{Fig8} (a) Tunneling spectra along a wire-isopotential line (the red line in Fig.~\ref{Fig7}a), at $B$ = 1~T. (b) and (c) Close-up views for the squared regions in panel (a). (d)-(f), Logarithmic color scale plots corresponding to panels (a)-(c), respectively. A zero-bias peak runs through the entire gate range. Greens lines in panels (e) and (f) are from the effective model fitting, which gives $\eta\approx$0.4 (or $q \approx 0.8$) for the sub-gap state.}
\end{figure}

The zero energy mode in Fig.~\ref{Fig7} with oscillating splittings indicates it is coupled Majorana modes. To estimate the nonlocality of the sub-gap states, we performed the quantum dot crossing experiment for this device, along a wire-isopotential [red arrow in Fig.~\ref{Fig7}(a)]. As shown in Fig.~\ref{Fig8}, the zero-bias peak traverses the whole measured region [logarithmic color scale in Fig.~\ref{Fig8} (d) for a clear view], with some finite broadening/splitting at dot level crossing points. Fit to the theory mentioned above, we can estimate the nonlocality of the sub-gap state is about  $\eta\approx$0.4 (or $q \approx 0.8$). For the same device, the degree of the nonlocality in Fig.~\ref{Fig8} is different with the one in Fig.~\ref{Fig4}. This is because gate tuning can significantly change the potential profile and/or effective Fermi wave vector, and thus leads to changes of nonlocality.

\section{Nonlocality of a trivial ABS}

\begin{figure}
\centering \includegraphics[width=8.5cm]{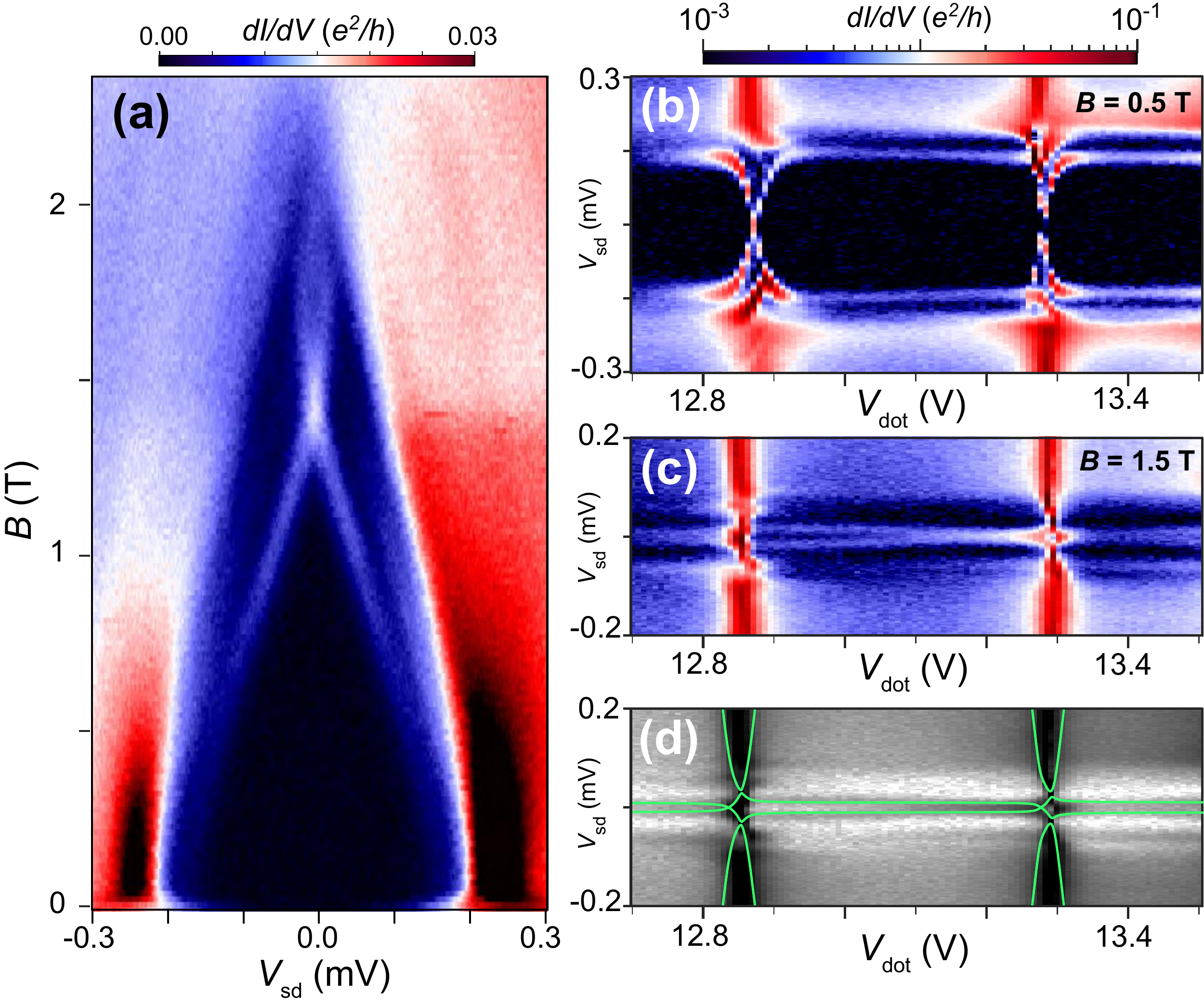}
\caption{\label{Fig9} (a) Tunneling conductance measurements as a function of magnetic field. A zero-bias peak develops around $B= 1.2$~T, splits and merges again at higher field. (b), (c) The dot-crossing tunneling spectra along a wire-isopotential direction at $0.5$, $1.5$~T, respectively. (d) Effective model fitting (green lines) for panel (d). The fitting gives $\eta\sim$ 0.85, roughly corresponding to $q \sim 0.28$.}
\end{figure}

We have also measured device C. In the magnetic field evolution of tunneling conductance for device C, a zero-bias peak emerges at $B=1.4$~T [Fig.~\ref{Fig9}(a)], which splits and forms a characteristic loop structure at higher field. To check the degree of nonlocality for the corresponding midgap states, dot-level crossing sweeps have been done at $B=0.5$~T and $B=1.5$~T [Fig.~\ref{Fig9}(b,c)]. It is evident that the zero-bias peak at $B=1.5$~T is sensitive to the dot levels. Fits to the model yield $\eta\sim$ 0.85, corresponding to $q \sim 0.28$, indicating that the midgap state is a rather spatially localized. The estimated weak nonlocality is consistent with the fact that the zero-bias peak is not robust in magnetic field evolution. The corresponding subgap state can be interpreted as MBSs with a large spatial overlap, or equivalently, trivial ABSs.

\section{Theoretical model and simulations}

The analysis and interpretation of the experimental data uses a theoretical description of the quantum dot-nanowire system following Ref.~\cite{Prada2017}. The resulting effective model captures the low-energy spectrum of the quantum dot-nanowire system as the dot states hybridize with Majorana bound states. Its ingredients are various properties of the corresponding wave functions, such as the spin orientation of dot and Majorana states, or the amplitude of the two Majoranas at the junction. 

The strategy is to phenomenologically model only the states involved in the anticrossings around zero energy, namely the low-lying dot states $d_{\uparrow,\downarrow}$ and the two Majorana bound states located at the junction ($\gamma_L$) and the far end ($\gamma_R$) of the nanowire (assuming they are spectrally separated from other states). Parameters of the effective model are the Majorana splitting $\delta$ and the hopping amplitudes $t_{L,R}$ from the quantum dot to the left and right Majoranas, respectively. Interestingly, this simple model is able to fully capture the transition from the trivial regime (where a local Andreev level would correspond to $t_L\sim t_R$ and arbitrary $\delta$) to the non-trivial topological regime, where nonlocal Majorana end states imply $\delta,t_R\ll t_L$. 

Other parameters are the quantum dot level $\epsilon_0$, the charging energy $U$, and the Majorana spin canting angles $\theta_L$ and $\theta_R$, of each Majorana wave function with respect to the spin quantization axis given by the Zeeman field $\mathcal{B}$. Using this minimal description, the Hamiltonian for a quantum dot interacting with the low-lying levels of the nanowire takes the form

\begin{widetext}
\begin{eqnarray}
H^\mathrm{eff}&=&\frac{1}{2}\left(d^\dagger_\uparrow,d^\dagger_\downarrow,d_\uparrow,d_\downarrow,\gamma_L,\gamma_R\right)\check{H}^\mathrm{eff}\left(d_\uparrow,d_\downarrow,d^\dagger_\uparrow,d^\dagger_\downarrow,\gamma_L,\gamma_R\right)^T, \nonumber\\
\frac{1}{2}\check{H}^\mathrm{eff}&=&\left(\begin{array}{cccccc}
\frac{\epsilon_0+\mathcal{B}+U\langle n_\downarrow\rangle}{2} & 0 & 0 & 0  &  t_L\sin\frac{\theta_L}{2} & -i t_R \sin\frac{\theta_R}{2}\\
0 & \frac{\epsilon_0-\mathcal{B}+U\langle n_\uparrow\rangle}{2}   & 0 & 0  & -t_L\cos\frac{\theta_L}{2} & -i t_R \cos\frac{\theta_R}{2}\\
0 & 0 & -\frac{\epsilon_0+\mathcal{B}+U\langle n_\downarrow\rangle}{2} & 0 & -t_L\sin\frac{\theta_L}{2} & -i t_R \sin\frac{\theta_R}{2}\\
0 & 0 & 0 & -\frac{\epsilon_0-\mathcal{B}+U\langle n_\uparrow\rangle}{2}   &  t_L\cos\frac{\theta_L}{2} & -i t_R \cos\frac{\theta_R}{2}\\
t_L\sin\frac{\theta_L}{2} & -t_L\cos\frac{\theta_L}{2} & -t_L\sin\frac{\theta_L}{2} & t_L\cos\frac{\theta_L}{2}     & 0  &  i\delta/2 \\
it_R\sin\frac{\theta_R}{2} & it_R\cos\frac{\theta_R}{2} & it_R\sin\frac{\theta_R}{2} & it_R\cos\frac{\theta_R}{2} & -i\delta/2 & 0
\end{array}\right).
\label{effmodel}
\end{eqnarray}
\end{widetext}

In the Coulomb blockade regime, dot occupancies are estimated from by their averages in the weak tunnelling limit $t_{L,R}\to 0$,
\begin{eqnarray}\label{meanfield}
\langle n_\downarrow\rangle&=&\langle n_\uparrow\rangle = 1 \space \textrm{ for $\epsilon_0<-U-\mathcal{B}$}, \nonumber\\
\langle n_\downarrow\rangle&=&1-\langle n_\uparrow\rangle = 1\textrm{ for $-U-\mathcal{B}<\epsilon_0<\mathcal{B}$}, \nonumber \\
\langle n_\downarrow\rangle&=&\langle n_\uparrow\rangle = 0 \space \textrm{ for $\mathcal{B}<\epsilon_0$}.
\end{eqnarray} 

The quantity $\eta=\sqrt{t_R/t_L}$ can be extracted from fitting transport spectroscopy measurements to the spectrum to Eq. \eqref{effmodel} using expressions for the energies involved in the avoided crossings,
\begin{eqnarray}
\epsilon_-^{M,D}&=&\sqrt{\frac{\delta^2}{2}+s_L^2+s_R^2\mp\sqrt{\left(\frac{\delta^2}{2}+s_L^2+s_R^2\right)^2-4s_L^2s_R^2}} \nonumber\\
\epsilon_+^{M,D}&=&\sqrt{\frac{\delta^2}{2}+c_L^2+c_R^2\mp\sqrt{\left(\frac{\delta^2}{2}+c_L^2+c_R^2\right)^2-4c_L^2c_R^2}}, \nonumber\\
\end{eqnarray}
where $s_i=2t_i\sin\frac{\theta_i}{2}$ and $c_i=2t_i\cos\frac{\theta_i}{2}$. By using these anaytical expressions to fit the data (see e.g. the green curves in Figs.~\ref{Fig4} of the main text), we can extract a good estimate of the degree of Majorana nonlocality. 

A value $t_R<t_L$, i.e. $\eta<1$ reflects a suppressed overlap of the two Majoranas, as demonstrated with direct comparison with a full microscopic model (see Section IV of Ref. \cite{Prada2017}). The limit $\eta=0$ corresponds to the case of zero overlap. In terms of the anticrossing structure, it was also shown that $\eta<1$ is associated to a spectral detachment between the Majorana-like and dot-like levels throughout the anticrossing. The perfectly nonlocal case $\eta=0$ manifests in unperturbed zero-energy Majorana states inside detached anticrossings, like those in Fig.~\ref{Fig4}.


\begin{thebibliography}{99}

\bibitem{Kitaev2000}
A. Y. Kitaev, \emph{Unpaired Majorana Fermions in Quantum Wires}, Physics-Uspekhi \textbf{131}, 130-136 (2001).

\bibitem{Read2000}
N. Read, and D. Green, \emph{Paired States of Fermions in Two Dimensions with Breaking of Parity and Time-reversal Symmetries and the Fractional Quantum Hall Effect}, Phys.~Rev.~B \textbf{61}, 10267 (2000).

\bibitem{Nayak2008}
C. Nayak, S. H. Simon, A. Stern, M, Freedman, and S. Das Sarma, \emph{Non-Abelian Anyons and Topological Quantum Computation}, Rev.~Mod.~Phys. \textbf{80}, 1083-1159 (2008).

\bibitem{DasSarmaNPJ2015}
S. Das Sarma, M. Freedman, and C. Nayak, \emph{Majorana Zero Modes and Topological Quantum Computation}, NJP Quantum Information \textbf{1}, 15001 (2015).

\bibitem{Moore1991}
G. Moore, and N. Read, \emph{Non-Abelions in the Fractional Quantum Hall Effect}, Nuclear Physics B \textbf{360}, 2362-396 (1991).

\bibitem{Fu2008}
L. Fu, and C. Kane, \emph{Superconducting Proximity Effect and Majorana Fermions at the Surface of a Topological Insulator}, Phys.~Rev.~Lett. \textbf{100}, 096407 (2008).

\bibitem{Goldstein2011}
G.~Goldstein and C.~Chamon, \emph{Decay Rates for Topological Memories Encoded with Majorana Fermions}, Phys.~Rev.~B \textbf{84}, 205109 (2011).

\bibitem{Budich2012}
J. C. Budich, S. Walter, and B. Trauzettel, \emph{Failure of Protection of Majorana Based Qubits Against Decoherence}, Phys.~Rev.~B \textbf{85}, 121405 (2012).

\bibitem{Lutchyn2010}
R. Lutchyn, J. D. Sau, and S. Das Sarma, \emph{Majorana Fermions and a Topological Phase Transition in Semiconductor-Superconductor Heterostructures}, Phys.~Rev.~Lett. \textbf{105}, 077001 (2010).

\bibitem{Oreg2010}
Y. Oreg, G. Refael, and F. von Oppen, \emph{Helical Liquids and Majorana Bound States in Quantum Wires}, Phys.~Rev.~Lett. \textbf{105}, 177002 (2010).

\bibitem{Lutchyn2017}
R. M. Lutchyn, E. P. A. M. Bakkers, L. P. Kouwenhoven, P. Krogstrup, C. M. Marcus, and Y. Oreg, \emph{Majorana zero modes in superconductor-semiconductor heterostructures}, Nature Reviews Materials, \textbf{105}, 52 (2018).

\bibitem{Aguado2017}
R. Aguado, \emph{Majorana quasiparticles in condensed matter}, Riv. Del Nuovo Cim. \textbf{40}, 523 (2017).

\bibitem{Mourik2012}
V. Mourik, K. Zuo, S. M. Frolov, S. R. Plissard, E. P. A. M. Bakkers, and L. P. Kouwenhoven, \emph{Signatures of Majorana Fermions in Hybrid Superconductor-Semiconductor Nanowire Devices}, Science \textbf{336}, 1003-1007 (2012).

\bibitem{Deng2012}
M. T. Deng, C. L. Yu, G. Y. Huang, M. Larsson, P. Caroff, and H. Q. Xu, \emph{Anomalous Zero-Bias Conductance Peak in a Nb-InSb Nanowire-Nb Hybrid Device}, Nano Lett. \textbf{12}, 6414-6419 (2012).

\bibitem{Das2012}
A. Das, Y. Ronen, Y. Most, Y. Oreg, M. Heiblum, and H. Shtrikman, \emph{Zero-bias Peaks and Splitting in an Al-InAs Nanowire Topological Superconductor as a Signature of Majorana Fermions}, Nat. Phys. \textbf{8}, 887-895 (2012).

\bibitem{Churchill2013}
H. O. H. Churchill, V. Fatemi, K. Grove-Rasmussen, M. T. Deng, P. Caroff, H. Q. Xu, and C. M. Marcus, \emph{Superconductor-nanowire Devices from Tunneling to the Multichannel Regime: Zero-bias Oscillations and Magnetoconductance Crossover}, Phys.~Rev.~B \textbf{87}, 241401 (2013).

\bibitem{Albrecht2016}
S. M. Albrecht, A. P. Higginbotham, M. Madsen, F. Kuemmeth, T. S. Jespersen, J. Nyg\aa rd, P. Krogstrup, and C. M. Marcus, \emph{Exponential Protection of Zero Modes in Majorana Islands}, Nature (London) \textbf{531}, 7593 (2016).

\bibitem{Deng2016}
M. T. Deng, S. Vaitiek\.{e}nas, E. B. Hansen, J. Danon, M. Leijnse, K. Flensberg, J. Nyg\aa rd, P. Krogstrup, and C. M. Marcus, \emph{Majorana Bound State in a Coupled Quantum-dot Hybrid-nanowire System}, Science, \textbf{354}, 6319 (2016). 

\bibitem{Chen2017}
J. Chen, P. Yu, J. Stenger, M. Hocevar, D. Car, S. R. Plissard, E. P. A. M. Bakkers, T. D. Stanescu, and S. M. Frolov, \emph{Experimental phase diagram of zero-bias conductance peaks in superconductor/semiconductor nanowire devices}, Sci. Adv. \textbf{3}, e1701476 (2017).

\bibitem{Zhang2017}
H. Zhang, C.-X. Liu, S. Gazibegovic, D. Xu, J. A. Logan, G. Wang, N. van Loo, J. D. S. Bommer, M. W. A. de Moor, D. Car, R. L. M. O. het Veld, P. J. van Veldhoven, S. Koelling, M. A. Verheijen, M. Pendharkar, D. J. Pennachio, B. Shojaei, J. S. Lee, C. J. Palmstr{\o}m, E. P. A. M. Bakkers, S. Das Sarma, and L. P. Kouwenhoven,  \emph{Quantized Majorana Conductance}, Nature, doi:10.1038/nature26142 (2018).

\bibitem{Perge2014}
S. Nadj-Perge, I. K. Drozdov, J. Li, H. Chen, S. Jeon, J. Seo, A. H. MacDonald, B. A. Bernevig, and A. Yazdani, \emph{Observation of Majorana Fermions in Ferromagnetic Atomic Chains on a Superconductor}, Science \textbf{346}, 602 (2014).

\bibitem{Suominen2017}
H. J. Suominen, M. Kjaergaard, A. R. Hamilton, J. Shabani, C. J. Palmstr{\o}m, C. M. Marcus, and F. Nichele, \emph{Zero-Energy Modes from Coalescing Andreev States in a Two-Dimensional Semiconductor-Superconductor Hybrid Platform}, Phys. Rev. Lett. \textbf{119}, 176805 (2017). 

\bibitem{Nichele2017}
F. Nichele, A. C. C. Drachmann, A. M. Whiticar, E. C. T. O'Farrell, H. J. Suominen, A. Fornieri, T. Wang, G. C. Gardner, C. Thomas, A. T. Hatke, P. Krogstrup, M. J. Manfra, K. Flensberg, and C. M. Marcus, \emph{Scaling of Majorana Zero-Bias Conductance Peaks}, Phys.~Rev.~Lett. \textbf{119}, 136803 (2017).

\bibitem{Liu2017}
C-X. Liu, J. D. Sau, T. D. Stanescu, and S. Das~Sarma, \emph{Andreev Bound States versus Majorana Bound States in Quantum Dot-nanowire-superconductor Hybrid Structures: Trivial versus Topological Zero-bias Conductance Peaks}, Phys. Rev. B \textbf{96} 075161 (2017).


\bibitem{Moore2017}
C. Moore, T. D. Stanescu, and S. Tewari, \emph{Two Terminal Charge Tunneling: Disentangling Majorana Zero Modes from Partially Separated Andreev Bound States in Semiconductor-superconductor Heterostructures}, arXiv:1711.06256 (2017).

\bibitem{Li2015}
J. Li, T. Yu, H.-Q. Lin, and J. Q. You, \emph{Probing the Nonlocality of Majorana Fermions via Quantum Correlations}, Sci. Rep. \textbf{4}, 4930 (2015).

\bibitem{Sau2015}
J. D. Sau, B. Swingle, and S. Tewari, \emph{Proposal to Probe Quantum Non-locality of Majorana Fermions in Tunneling Experiments}, Phys. Rev. B \textbf{92}, 020511 (2015).

\bibitem{Rubbert2016}
S. Rubbert, and A. R. Akhmerov, \emph{Detecting Majorana Nonlocality Using Strongly Coupled Majorana Bound States}, Phys. Rev. B \textbf{94}, 115430 (2016).

\bibitem{Hell2017}
M. Hell, K. Flensberg, and M. Leijnse, \emph{Distinguishing Majorana Bound States from Localized Andreev Bound States by Interferometry}, arXiv:1710.05294 (2017).

\bibitem{Prada2017}
E. Prada, R. Aguado, and P. San-Jose, \emph{Measuring Majorana Non-locality and Spin Structure with a Quantum Dot}, Phys. Rev. B \textbf{96}, 085418 (2017).

\bibitem{Clarke2017}
D. J. Clarke, \emph{Experimentally Accessible Topological Quality Factor for Wires with Zero Energy Modes}, Phys. Rev. B \textbf{96}, 201109 (2017).

\bibitem{Ben-Shach2015}
 G. Ben-Shach, A. Haim, I. Appelbaum, Y. Oreg, A. Yacoby, and B. I. Halperin, \emph{Detecting Majorana modes in one-dimensional wires by charge sensing}, Phys. Rev. B \textbf{91}, 045403 (2015).
 
\bibitem{Pinning2017}
F. Dom\'{i}nguez, J. Cayao, P. San-Jose, R. Aguado, A. L. Yeyati, and E. Prada, \emph{Zero-energy Pinning from Interactions in Majorana Nanowires}, Npj Quantum Materials, \textbf{2}, 13 (2017). 

\bibitem{Deng2014}
M. T. Deng, C. L. Yu, G. Y. Huang, M. Larsson, P. Caroff, and H. Q. Xu, \emph{Parity independence of the zero-bias conductance peak in a nanowire based topological superconductor-quantum dot hybrid device}, Sci. Rep. \textbf{4}, 7261 (2014).

\bibitem{Krogstrup2015}
P. Krogstrup, N. L. B. Ziino, W. Chang, S. M. Albrecht, M. H. Madsen, E. Johnson, J. Nyg\aa rd, C. M. Marcus, and T. S. Jespersen, \emph{Epitaxy of Semiconductor-superconductor Nanowires}, Nat.~Mat. \textbf{14}, 400-406 (2015).

\bibitem{Chang2015}
W. Chang, S. M. Albrecht, T. S. Jespersen, F. Kuemmeth, P. Krogstrup, J. Nyg\aa rd, and C. M. Marcus, \emph{Hard Gap in Epitaxial Semiconductor-superconductor Nanowires}, Nat.~Nanotechnol. \textbf{10}, 232-236 (2015).

\bibitem{MBSgfactor}
S. Vaitiek\.{e}nas, M. T. Deng, J. Nyg\aa rd, P. Krogstrup, and C. M. Marcus, \emph{Effective g-factor in Majorana Wires}, arXiv:1710.04300 (2017).

\bibitem{Antipov2018}
A. E. Antipov, A. Bargerbos, G. W. Winkler, B. Bauer, E. Rossi, and R. M. Lutchyn, \emph{Effects of Gate-induced Electric Fields on Semiconductor Majorana Nanowires}, arXiv:1801.02616 (2018).

\bibitem{Mikkelsen2018}
A. E. G. Mikkelsen, P. Kotetes, P. Krogstrup, and K. Flensberg, \emph{Hybridization at Superconductor-semiconductor Interfaces in Majorana Devices}, arXiv:1801.03439 (2018).

\bibitem{Reeg2018}
C. Reeg, D. Loss, and J. Klinovaja, \emph{Metallization of Rashba Wire by Superconducting Layer in the Strong-proximity Regime}, arXiv:1801.06509 (2018).

\bibitem{Lee2013}
E. J. H. Lee, X. Jiang, M. Houzet, R. Aguado, C. M. Lieber, and S. De Franceschi, \emph{Spin-resolved Andreev Levels and Parity Crossings in Hybrid Semiconductor-superconductor Nanostructures}, Nat.~Nanotechnol. \textbf{9}, 79-84 (2013).

\bibitem{Pillet2010}
J.-D. Pillet, C. H. L. Quay, P. Morfin, C. Bena, a. L. Yeyati, and P. Joyez, \emph{Andreev Bound States in Supercurrent-carrying Carbon Nanotubes Revealed}, Nat. Phys. \textbf{6}, 965-969 (2010).

\bibitem{Mason2011}
T. Dirks, T. L. Hughes, S. Lal, B. Uchoa, Y.-F. Chen, C. Chialvo, P. M. Goldbart, and N. Mason, \emph{Transport through Andreev Bound States in a Graphene Quantum Dot}, Nat. Phys. \textbf{7}, 386-390 (2011).

\bibitem{Chang2013}
W. Chang, V. E. Manucharyan, T. S. Jespersen, J. Nyg\aa rd, and C. M. Marcus, \emph{Tunneling Spectroscopy of Quasiparticle Bound States in a Spinful Josephson Junction}, Phys. Rev. Lett. \textbf{110}, 217005 (2013).

\bibitem{Chevallier2018}
D. Chevallier, P. Szumniak, S. Hoffman, D. Loss, and J. Klinovaja, \emph{Topological Phase Detection in Rashba Nanowires with a Quantum dot}. Phys. Rev. B \textbf{97}, 45404 (2018).

\bibitem{Mishmash2017}
R. V. Mishmash, D. Aasen, A. P. Higginbotham, and J. Alicea, \emph{Approaching a Topological Phase Transition in Majorana Nanowires}, Phys. Rev. B \textbf{93}, 245404 (2016).




\end{thebibliography}

\end{document}